\documentclass{aastex63}

\usepackage{longtable}
\usepackage{tablefootnote}

\usepackage{xcolor}

\begin{document}
\title{Sunquakes of Solar Cycle 24}
\correspondingauthor{Ivan N. Sharykin}
\email{ivan.sharykin@phystech.edu}

\author{Ivan N. Sharykin}
\affiliation{Space Research Institute (IKI) of the Russian Academy of Sciences, Profsoyuznaya 84/32, Moscow, 117997, Russia}
\affiliation{Moscow Institute of Physics and Technology, Institutsky lane 9, Dolgoprudny, Moscow region, 141700, Russia}
\author{Alexander G. Kosovichev}
\affiliation{Department of Physics, New Jersey Institute of Technology, University Heights, Newark, NJ 07102}
\affiliation{Center for Computational Heliophysics, New Jersey Institute of Technology, University Heights, Newark, NJ 07102}
	

\keywords{sunquakes --- helioseismology --- flares}


\begin{abstract}

The paper presents results of a search for helioseismic events (sunquakes)  produced by M-X class solar flares during Solar Cycle 24. The search is performed by analyzing photospheric Dopplergrams from Helioseismic Magnetic Imager (HMI). Among the total  number of 500 M–X class  flares,  94 helioseismic events  were detected. Our analysis  has shown that many strong sunquakes were produced by solar flares of low M class (M1-M5), while in some powerful X-class flares helioseismic waves were not observed or were weak. Our study also revealed that only several active regions were characterized by the most efficient generation of helioseismic waves during flares. We found that  the sunquake  power correlates  with the maximum  value  of the soft X-ray flux time derivative better  than with the X-ray class,  indicating  that  the sunquake mechanism is associated  with high-energy particles.  We also show that the seismically  active  flares  are more impulsive  than  the flares  without   helioseismic perturbations. We present a new catalog of helioseismic solar flares, which opens opportunities  for performing statistical studies to better understand the physics of sunquakes as well as the flare energy release and transport.
\end{abstract}



\section{Introduction}

Strong photospheric perturbations during solar flares are believed to be accompanied by generation of helioseismic waves, also referred as ``sunquakes''. This phenomenon was initially suggested by \cite{Wolff1972}, theoretically predicted by \cite{Kosovichev1995}, and later discovered by \cite{Kosovichev1998} using Dopplergrams from Michelson Doppler Imager (MDI) onboard Solar Orbital Heliospheric Observatory (SOHO)\citep{Scherrer1995}. Helioseismic waves are observed in photospheric line-of-sight (LOS) Dopplergrams as concentric (usually highly anisotropic) waves spreading out from an initial photospheric flare impacts observed during the impulsive phase. They represent  acoustic waves  that  travel  through the solar  convective  zone, where the corresponding  acoustic rays are curved due to refraction  caused by the increasing with depth temperature. The acoustic waves  emitted into the solar interior are reflected back to the solar surface,  and observed as traveling circular wave  packets in the Dopplergrams. Helioseismic events  are usually  associated with compact  photospheric  perturbation  and appearance  of continuum emission close to the acoustic sources reconstructed by the helioseismic holography method \citep[][]{Lindsey1997,Donea1999,Lindsey2000}, see also statistical work of \citet{Buitrago-Casas2015}.

Observational properties and theories of sunquakes can be found in the reviews of \cite{Donea2011} and \cite{Kosovichev2015}. We briefly mention hypotheses  of sunquake  generation.  In particular, the mechanism of the flare-excited helioseismic waves had been considered in the frame of a beam-driven hypothesis. It assumes that the initial perturbation is produced by the hydrodynamic response to accelerated electrons injected into the chromosphere  \citep{Kosovichev1995}, which is confirmed by the close temporal  and spatial  association  of sunquake  sources with hard X-ray emission sources \citep[e.g.][]{Kosovichev2006, Kosovichev2007, Sharykin2017}.  The backwarming effect of photospheric heating by flare UV radiation may also induce pressure perturbation needed to generate sunquake waves \cite{Donea2011}. In addition, the plasma momentum can be transferred by a sharp enhancement of pressure gradient due to eruption of a magnetic flux-rope \citep[e.g.][]{Zharkov2011,Zharkov2013} or by an impulse Lorentz force which can be stimulated by changing magnetic fields in the lower solar atmosphere \citep{Hudson2008,Fisher2012,AlvaradoGomez2012, Burtseva2015,Russell2016}. \citet{Sharykin2015a} and \citet{Sharykin2015b} discussed that rapid dissipation of electric currents in the low atmosphere could also explain sunquake initiation. The exact cause of sunquakes is still uknown.

Perhaps,  different sunquake events  can be caused  by  different  mechanisms,  or, different  mechanisms  can operate together.   In order to understand  why some flares produce sunquakes  and some do not, and what physical  properties  lead to the flare seismic activity, it is important  to perform statistical studies.  Initial catalogs  of sunquakes  were presented  by  \citet{Buitrago-Casas2015} and \citet{Besliu-Ionescu2017} for Solar Cycles 23 and 24 (up to February of 2014). These surveys  reported 23 and 18 helioseismic events  correspondingly and the first work considered only flares with pronounced hard X-ray emission above 50 keV according to RHESSI (Reuven Ramaty High Energy Solar Spectroscopic Imager, \citet{Lin2002}) catalogue. More recently, \citet{Chen2019} analyzed 60 strong flares (with the GOES X-ray class greater than M5) in Solar Cycle 24 to search for sunquakes by a helioseismic holography method. A total of 24 flares were found to be seismically active, giving a total of 41 sunquakes.  It is worth noting that  analyses  of flares during Cycle 23 using MDI data  found sunquake  events  only for X-class flares while analysis  of Cycle 24 data  from HMI data  discovered  sunquakes for M-class and even a C-class flare \citep{Sharykin2015a}.

In order to obtain a clear understanding of the mechanism of sunquakes and answer the questions: why some flares produce sunquakes and other do not, and how the sunquake power depends on flare properties, it is necessary to carry out statistical studies, the first step of which is to create a comprehensive catalog of the sunquake events.  The scope of this work is to perform  a search of sunquake events in all M-X class solar flares observed on the solar disk during Solar Cycle 24. We perform the following tasks: 

\begin{enumerate}
\item Develop a comprehensive catalog of sunquake events during Solar Cycle 24.
\item Compare helioseismic flares and flares without photospheric disturbances from the point of view of basic parameters which can be extracted from the GOES soft X-ray data \citep{Bornmann1996}, e.g.,  X-ray class, duration of the impulsive phase and maximum value of the X-ray flux time derivative.
\item Investigate the relationship between the sunquake energy estimated from the acoustic holography technique and the flare X-ray properties from the GOES data.
\end{enumerate}

The X-ray characteristics in these tasks provide the most important parameters of the flare energetics. A series of recent studies of the global energetics of flares and CMEs has found the following overall average ratios: $51\% \pm 17\%$ for electron acceleration, $17\% \pm 17\%$ for ion acceleration, $7\% \pm 14\%$ for CME kinetic energies, and $7\% \pm 17\%$ for direct heating \citep{Aschwanden2019}. The time derivative of the soft X-ray (SXR) flux is correlated with the hard X-ray (HXR) flux (the Neupert effect) \citep{Neupert1968, Dennis1993}.  It is a result of electron beam heating of the chromosphere, where the non-thermal HXR emission is produced by the electron beam \citep{brown1971}. The beam heats the chromospheric plasma to coronal temperatures causing its evaporation and increase of the SXR emission in the corona, as follows from the radiative hydrodynamics simulations \citep[e.g.][]{Livshits1981,Fisher1985,Kosovichev1986,Allred2015}.  Recent helioseismic modeling of sunquakes by \citet{Stefan2019} showed that at least half the studied events were consistent with the electron beam hypothesis. Therefore, the statistical comparison of the X-ray and helioseismic energetics is an important step to understanding physical processes associated with the flare helioseismic response.

The paper is divided into four sections. The first one describes methods to identify flare helioseismic waves and their sources from the SDO/HMI data. The second section describes a catalog of sunquakes of Solar Cycle 24, and presents a summary of the most seismically active regions. A statistical study of the seismic and non-seismic flares is presented in the third section. The last section summarizes results and formulates conclusions.
                                                      
\section{Identification of Sunquakes and Their Sources} 

\begin{figure}[ht]
\centering
\includegraphics[width=1.0\linewidth]{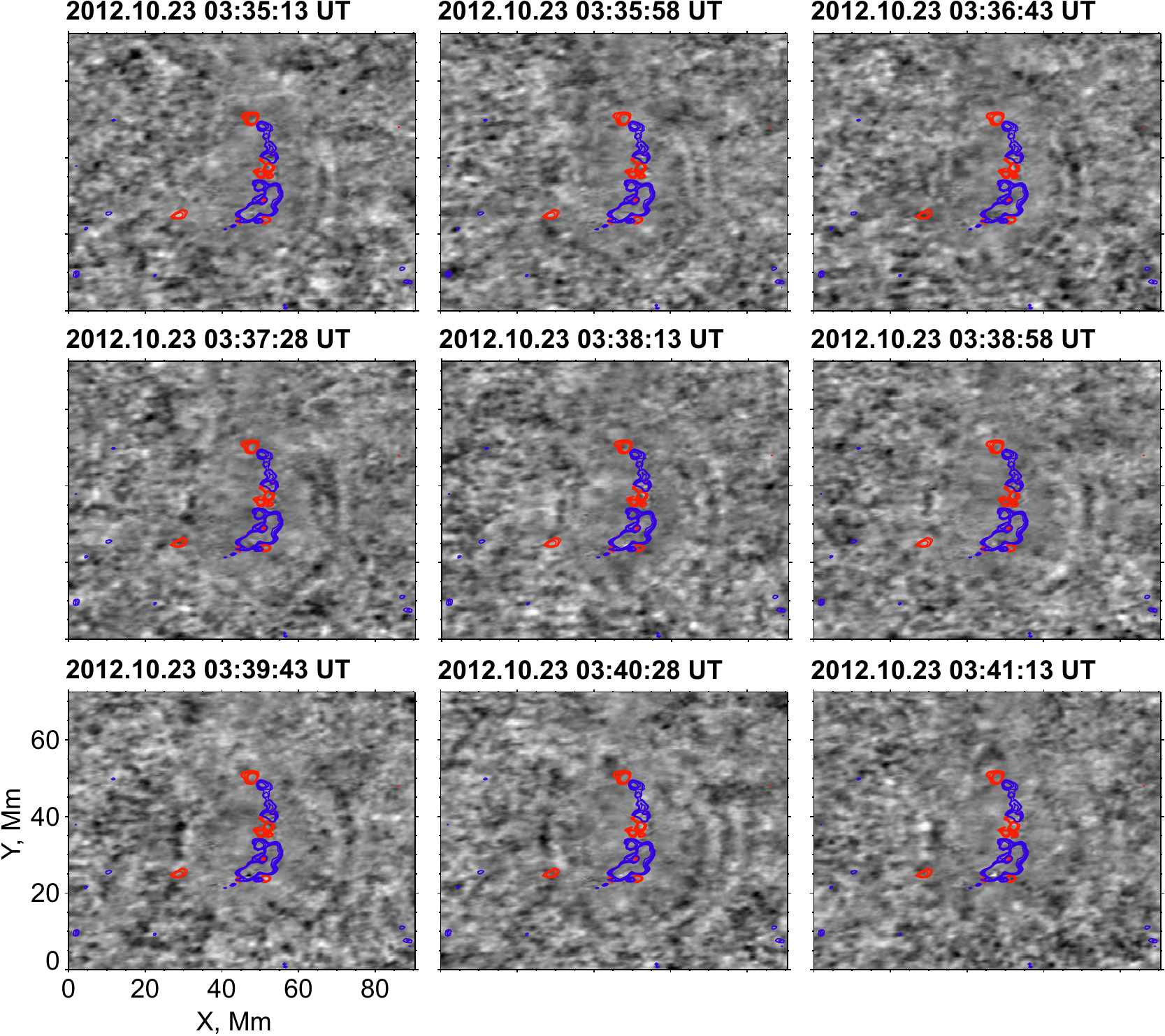}
\caption{A sequence of the HMI Dopplergram running differences filtered in the frequency range of 5-7 mHz for the sunquake event associated with X1.8 flare of October 23, 2012. Red and blue contours highlight regions of strong upward and downward velocity perturbations with the absolute value higher than 1~km/s.}
\label{SQwaves}
\end{figure}

For analysis, we use the HMI line-of-sight (LOS) Dopplergrams with 45-sec cadence  and 1 arcsec spatial resolution (0.5~arcsec/px) \citep{Scherrer2012}. To search for helioseismic waves and their sources we used time sequences of the running difference of derotated (removed effects of solar rotation) Dopplergrams remapped onto the heliographic grid (Fig.\,\ref{SQwaves}). To isolate the wave signal from convective noise we applied to each pixel of the Dopplergram data cubes a Gaussian frequency filter with a central frequency of 6 mHz and standard deviation width of 2 mHz, which is optimal for rendering the sunquake signal \citep{Donea1999}.


For the search of sunquakes, we consider only solar flares that produced perturbation in the photospheric HMI Dopplergrams. This is justified because sunquakes and their sources appear as photospheric perturbations. 
First, we visually inspected all Dopplergram series to find Doppler impacts during flares.  It is worth noting that we used subjective visual inspection due to the following reason. The initial helioseismic impact usually starts from strong photospheric Doppler perturbations seen during the flare impulsive phase. However, it is very difficult to characterize these perturbations quantitatively (e.g. amplitude value) due to the Fe~I line shape distortions do not allow to reliably determine the LOS velocity associated with this impact \citep{RajaBayanna2014,Sun2017}. This can be done in the frame of a future separate work following the forward modeling approach \citep{Sadykov2019}. At this stage of the work (statistical survey of the flare seismicity), we include only events with the impacts during the flare impulsive phase, which can be visually identified in the HMI Dopplergrams above the noise level.
We generated running time-difference Doppler movies for all flares and found 181 events with the Doppler impacts of different magnitudes. Then, we used three approaches to find helioseismic waves in the selected flares:
\begin{enumerate}

\item Create movies showing time sequences of running differences of derotated HMI Dopplergrams projected onto the heliographic grid and filtered in the frequency range of 5-7 mHz. The helioseismic waves are detected by visual inspection of these movies.

\item Select photospheric impacts detected in the HMI Dopplergrams (also derotated, reprojected and filtered), and construct time-distance (TD) diagrams. The helioseismic waves are detected in the form of a characteristic ridge pattern in the TD diagrams \citep{Kosovichev1998}.

\item Reconstruct the two-dimensional structure of seismic sources by using the helioseismic holography method \citep{Lindsey2000}. This approach employs a theoretical Green function of helioseismic waves to calculate the egression acoustic power corresponding to the observed Doppler velocity perturbations.

\end{enumerate}                                                                    

The most direct way to find sunquake waves is visual inspection of the Dopplergrams movies (the SQ movie method). Generally, the sunquake ripples appear about 10-15 min after the initial impact. The excited helioseismic waves travel beneath the solar surface before they appear on the surface in the form of sunquake ripples. The travel distance and travel times are shorter for the waves with shorter horizontal wavelengths. However, such waves in the sunquake signal are strongly damped \citep{Stefan2019}, In addition, the sunquake ripples usually are not observed when travel in strong magnetic field regions (sunspot umbra and penumbra) due to the wave absorption \citep{Zhao2016} and reflection in deeper subsurface layers \citep{Zhao2012}. While the visual inspection provides the most unambiguous detection in the case of strong sunquakes, and allows estimating the wave anisotropy and tracking propagation through the active region, it becomes subjective in the case of weak events. The problem is that in addition to the random oscillation background, there are many sporadic weak acoustic sources which can be misinterpreted as sunquakes. We identify only unambiguous events as  helioseismic when circular-shape wave packets are spreading out from initial photospheric impacts, as shown Fig.~\ref{SQwaves} (similarly to water ripples when a pebble is dropped into water). Such wave packets are usually well seen in the active region areas where acoustic background amplitude is suppressed due to magnetic field. 

\begin{figure}[ht]
\centering
\includegraphics[width=0.95\linewidth]{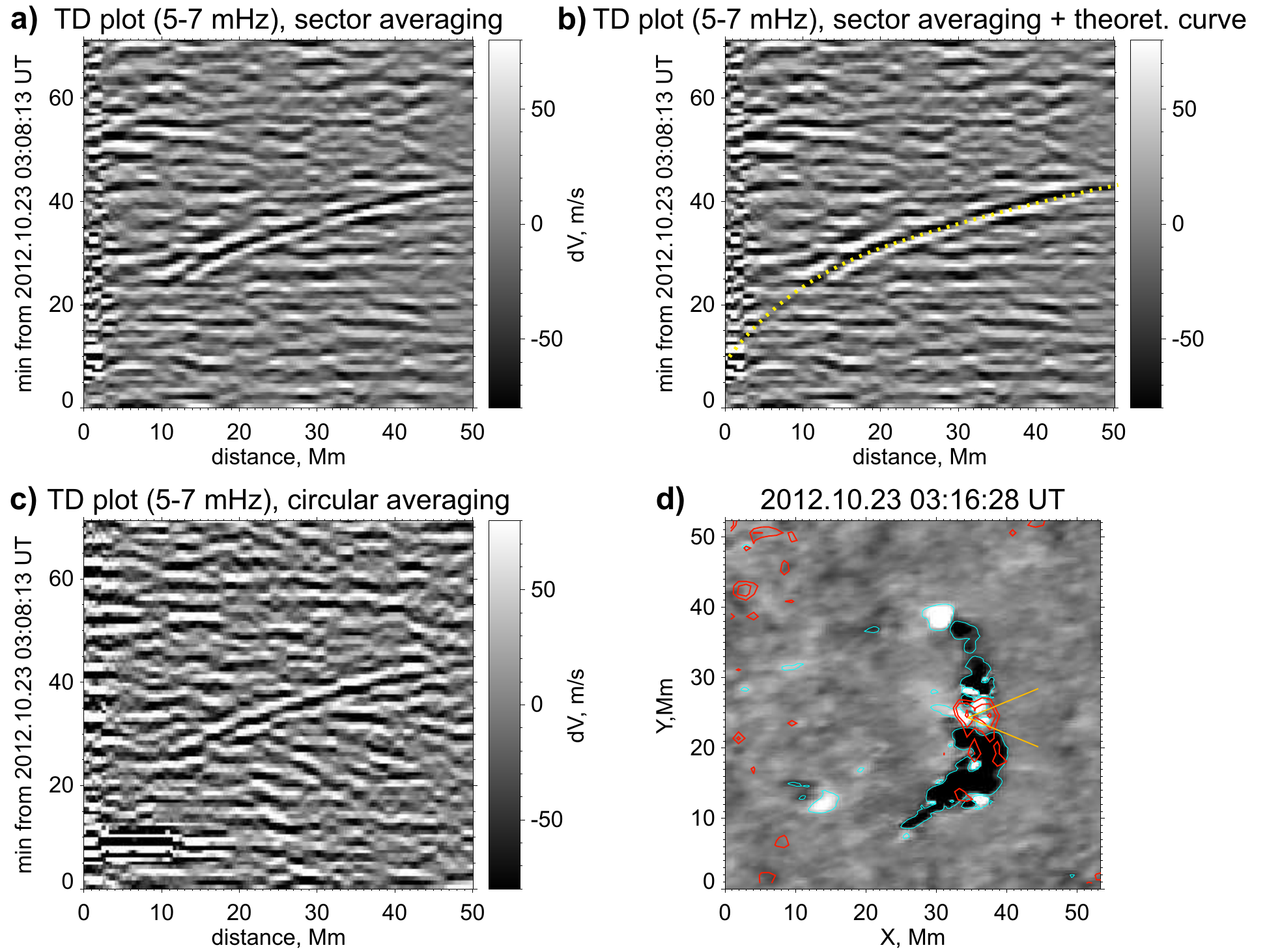}
\caption{Time-distance (TD) diagrams in the frequency range of 5-7 mHz for the sunquake event associated with X1.8 flare of October 23, 2012. Panels (a) and (b) show the TD diagram for the case of sector averaging around a point selected in the region of a strong photospheric perturbation and the acoustic source deduced by the acoustic holography method. The sector and selected source point are shown by yellow lines in panel (d). Yellow dotted line in panel (b) corresponds to the ray-theoretical prediction of the helioseismic wave path. Panel~(c) shows the TD diagram obtained from Dopplergram averaging over the full circle. Panel~(d) presents the photospheric Dopplergram difference map (white-black background image) filtered in the frequency range of 5-7 mHz with overplotted red contours (30, 50 and 90\% levels relative to maximum value) highlighting the acoustic sources deduced from acoustic holography. Cyan contours show regions with the absolute value of velocity perturbation higher than 200 m/s.}
\label{TDplot1}
\end{figure}

The other two methods, time-distance and acoustic holography, are less subjective, but depend on the model of the solar interior structure and theory of helioseismic waves. The  helioseismic waves can be identified as a characteristic ridge in  the time-distance (TD) diagram that shows the wave signal averaged for given distances around a reference (wave source) point. This point is selected in the area of the initial photospheric flare disturbances and the strongest acoustic sources deduced by the acoustic holography. We construct two types of the TD diagrams. The first one is obtained by circular averaging. In this case, for each time moment (on the time axis of the TD diagram) we calculate a one-dimensional distribution of the Dopplergram signal obtained by averaging it along circles with radii equal to the distances (on the distance axis of the TD diagram) from 0 to 50 Mm. Another type of the TD diagram, which is used for highly anisotropic wave fronts, is obtained by averaging the signal over sectors. The sectoral TD diagrams are calculated for the angular range covering $\pi/4$ for distances from 0  to 50 Mm. The sector direction is selected to find the strongest helioseismic signal with a step of $\pi/8$ (so we investigate 16 TD diagrams for each event). To make conclusion that the observed wave pattern in the TD diagram corresponds to a sunquake event we compare it with the theoretical time-distance relation calculated in the ray approximation for a standard solar interior model \citep{Christensen-Dalsgaard1993}. It is marked by dashed curve in the TD~diagrams in Fig.\,\ref{TDplot1}. The position of the wave ripples in the TD diagram fits to the theoretical model. Thus, the observed wave was generated in the source corresponding to the Dopplergram disturbance around the selected reference point.

\begin{figure}[ht]
\centering                                                         \includegraphics[width=0.85\linewidth]{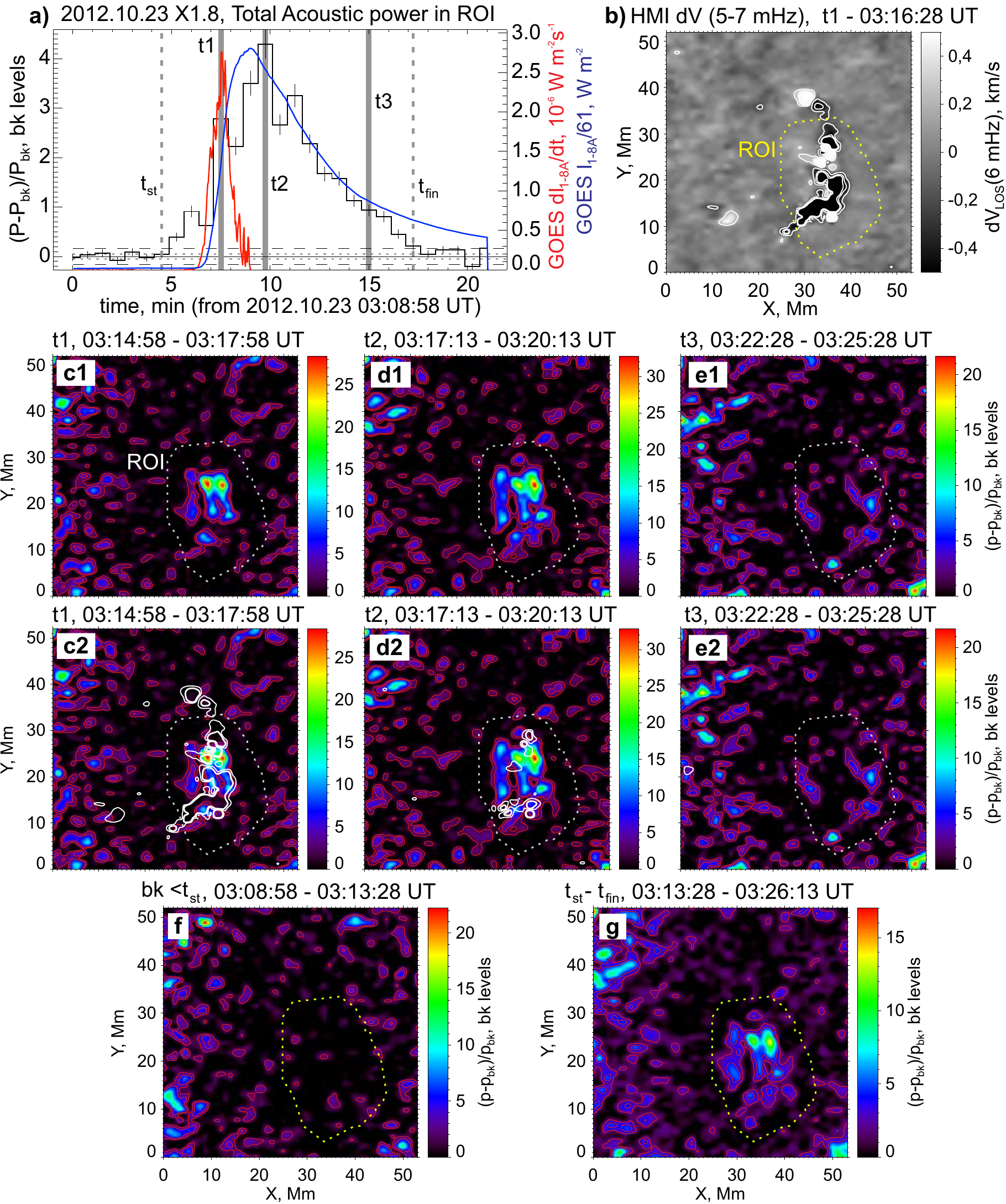}
\caption{Evolution of sunquake impacts investigated by acoustic holography in the frequency range of 5-7 mHz for X1.8 flare of October 23, 2012. Black histogram in panel~(a) shows the total acoustic power flux (in the units of background flux) within the region of interest (ROI) shown by dashed contours in other panels. The sunquake signal is marked between two vertical lines with labels $t_{st}$ and $t_{fin}$. Blue and red curves correspond to GOES 1-8~\AA{} flux and its time derivative. For visualization purpose the GOES flux is divided by a factor written in the axis title to match to maximum value of its time derivative. Panel (b) shows the photospheric Dopplergram difference map (white-black background image) filtered in the frequency range of 5-7 mHz for time moment t1 (first vertical gray line in panel~a).  Three gray vertical lines (marked as t1-3) in panel~(a) correspond to three time moments selected to demonstrate the acoustic sources in the holography images shown in panels~(c)-(e). Panels with labels 1 and 2 show the same acoustic egression maps. Contours in panels~(c2)-(e2) highlight regions with the absolute value of velocity perturbation higher than 200~m/s. Bottom panels~(f) and~(g) show  the integrated (sunquake signal between $t_{st}$ and $t_{fin}$) and background (signal before $t_{st}$) acoustic egression map, respectively.}
\label{Egress1}
\end{figure}

The helioseismic holography method \citep{Lindsey1997,Donea1999,Lindsey2000} is based on the idea of using a theoretical model of helioseismic waves to reconstruct the two-dimensional distribution of acoustic sources. This approach uses a theoretical Green function of helioseismic waves for the standard solar interior model to calculate the egression acoustic power corresponding to the Doppler velocity perturbations. An example of the egression acoustic power map calculated in the frequency range of 5-7 mHz is shown in Fig.\,\ref{Egress1}. We calculated this map by summing up the egression acoustic power snapshots within a time interval found from the uncertainty principle: $\Delta t\sim 1/\Delta \nu\approx 500$ sec, where $\Delta \nu=$2 mHz. This time interval corresponds to the appearance of strong Doppler velocity perturbations. The egression power map is compared with the corresponding Dopplergram.

\begin{figure}[ht]
\centering                                                          \includegraphics[width=\linewidth]{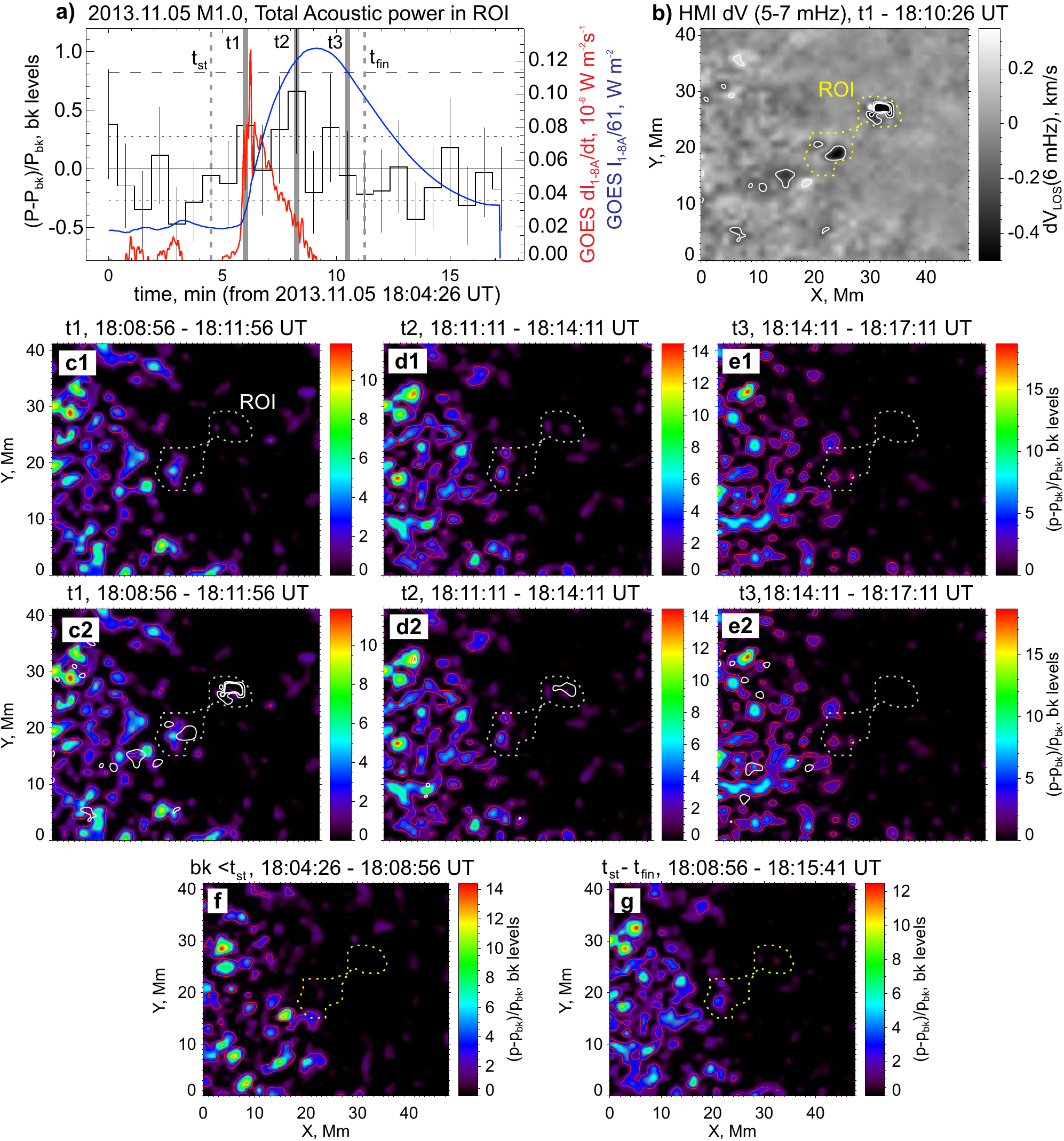}
\caption{Evolution of sunquake impacts investigated by acoustic holography in the frequency range of 5-7 mHz for M1.0 of November 5, 2013. Panels (a)-(g) show the same properties as in Fig.~\ref{Egress1}.}
\label{Egress2}
\end{figure}

Figures~\ref{Egress1} and~\ref{Egress2} illustrate the acoustic holography method applied to two different flares. In these figures, we present the acoustic egression power maps constructed from five 45-sec snapshots of the 225~sec duration, which is approximately equal to $\Delta t/2= 1/(2\Delta \nu) =250$~sec. Figures~\ref{Egress1}  presents a very strong sunquake event during X1.8 solar flare of October 23, 2012. This event is an example of the flare with an obvious helioseismic impact seen in the egression power maps. In this case, all three methods provide a robust detection: the Dopplergram movie revealed very clear wave ripples, and the time-distance analysis (see more details in the text below) confirmed that these waves were acoustic and spread from the initial photospheric perturbations. We show the photospheric flare impacts for one time moment in the corresponding frequency-filtered Dopplergram time-difference in Fig.~\ref{Egress1}b and, also, indicate them for three time moments by white contours overlaid on the acoustic power maps in panels (c2)-(e2). 

The advantage of the helioseismic holography technique is that it provides a quantitative estimation of the strength of the sunquake source. After analysis of many helioseismic events we decided to use the following criteria for detecting sunquakes by this technique. First of all, the reconstructed acoustic sources have to be in the area of the initial photospheric perturbations, usually observed during the flare impulsive phases. Other important criteria are the magnitude and timing of the total helioseismic signal. If the acoustic power spatially integrated over the area that includes the photospheric perturbations and acoustic sources (shown by dashed contours in Fig.~\ref{Egress1}b-g)  exceeds three background noise levels (calculated for a time period before appearance of the photospheric impacts), then we identify the signal as a helioseismic event. We also introduce a special class of helioseismic event candidates for the cases when weak acoustic sources are observed around photospheric perturbations in the Dopplergrams, but the calculated acoustic power is below the three sigma background level. Such weak event observed during M1.0 flare of November 5, 2013, is demonstrated in Fig.\,\ref{Egress2}. In this case, the holography method does not provide a clear detection. It is worth noting that the existence of sunquake waves in this flare was verified by the SQ movies and the time-distance diagram. We identified the holography signals as sunquake candidates only in the case of non-detection of waves in the movie and in the time-distance plot.

\begin{figure}[ht]
\centering
\includegraphics[width=0.95\linewidth]{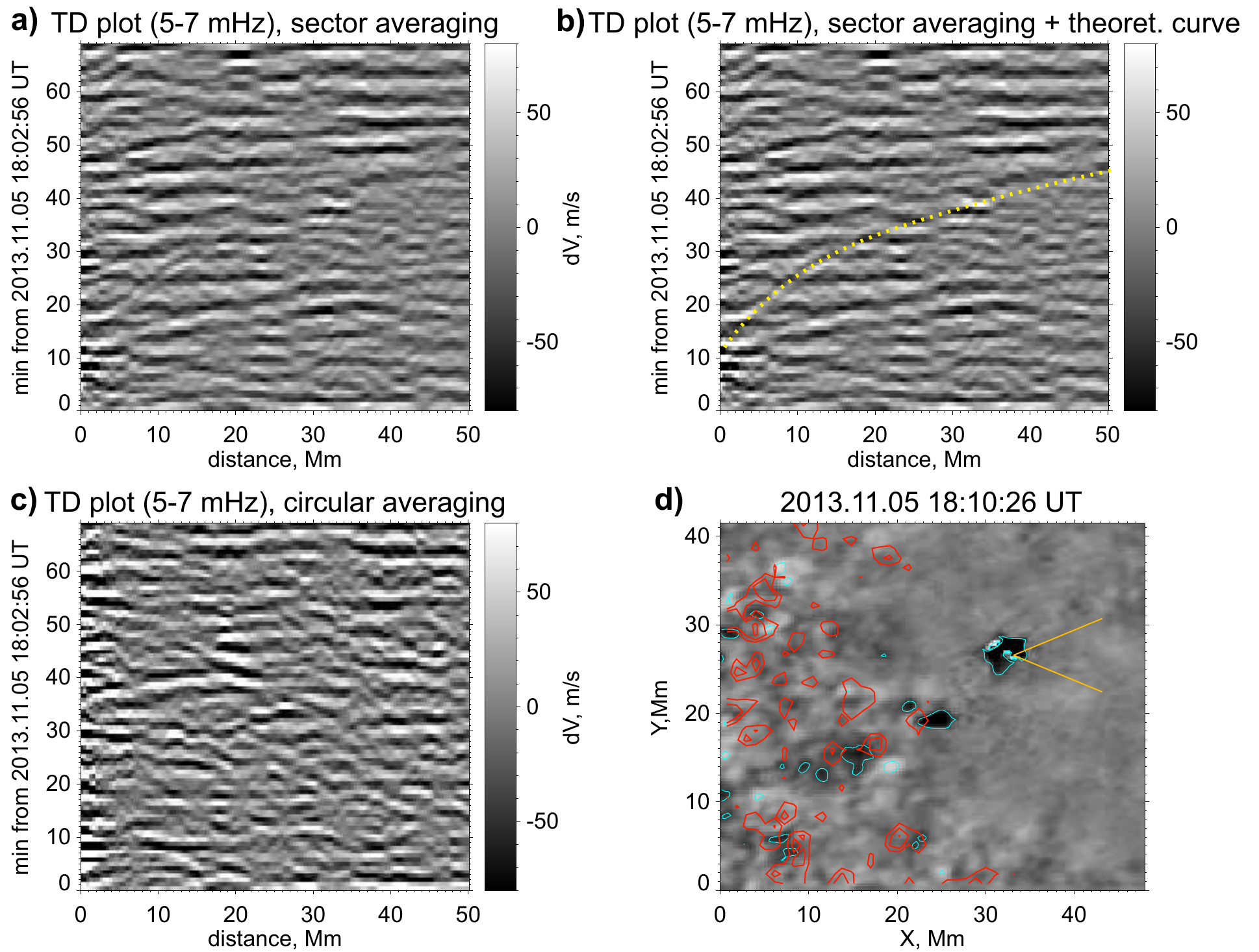}
\caption{ Time-distance diagrams in the frequency range of 5-7 mHz for sunquake event associated with M1.0 flare of November 5, 2013. This flare is an example of weak sunquake events found in the Dopplergram movie with weak wave signals seen in the TD diagram and the egression power map deduced from the acoustic holography. Panels (a)-(d) show the same properties as in Fig.~\ref{TDplot1}.}
\label{TDplot2}
\end{figure}

Figures~\ref{TDplot1} and~\ref{TDplot2} demonstrate examples of the TD diagrams for the flares for which the acoustic holography results were summarized in Figures~\ref{Egress1} and~\ref{Egress2}. For the X1.8 flare we have a clear wave signal in the TD diagrams for both types: the sectoral (a-b) and circular (c) averaging. Considering the weak M1 flare of November 5, 2013 in Fig.~\ref{TDplot2}, we found the sunquake signal only for the sector in the direction shown by yellow lines in panel (d).

All analysis methods described in this section have been employed to find sunquakes and possible candidates in all M-X class flares with photospheric perturbations. In the next section we present results of this search.

\section{Catalog of Sunquakes} 

Our results of searching for sunquakes among the M-X class flares with photospheric disturbances are presented in Table~\ref{table1}. The first four columns show the basic information about flares: Date, GOES class, NOAA number and angular position on the solar disk. The next three columns show results of identification of helioseismic waves. Plus or minus signs mean positive or negative identification using the three methods discussed in the previous section. The question marks indicate the potential candidates deduced from the Holography method.

Summary of the catalog shown in Table~\ref{table1} is the following:
\begin{enumerate}
\item Total number of the analyzed M and X class flares from 13-Feb-2011 to 8-Sep-2017: 507.
\item Total number of flares with photospheric perturbations: 181.
\item Total number of helioseismic events  registered by each of the three methods: 62 (Movie method), 80 (Holography method), 81 (TD method) ($+$29 Candidates deduced from Holography method)
\item Total number of sunquakes  detected by all three methods: 54.
\item Total number of sunquakes  detected by at least one method: 94 ($+$20 Candidates) $=$ 114 sunquakes. This means that more than half of the flares with photospheric perturbations are accompanied by helioseismic response.
\item Total number of Active Regions produced sunquakes: 35.
\item Eight of these Active Regions  produced more than 60\% of sunquakes  ($>5$ events/AR).
\end{enumerate}

Our statistical analysis confirmed complexity of the sunquake phenomenon discussed in the previous works briefly discussed in Introduction. First of all, we often observed sunquakes in relatively weak (low M class) events. Sometimes, the helioseismic sources in low M-class flares were more powerful than in X-class flares. Moreover, there were X-class flares without any manifestations of helioseismic response. We found many sunquakes during flares that occurred close to the solar limb: 31 and 9 events were at the angular distances from the disk center $>700$ and $>800$ arcsec, respectively. Previously, it was thought that it is hard to observe helioseismic waves close to the limb because of small amplitude of the LOS Doppler velocity variations due to the projection effect and foreshortening. 
It is worth noting that morphology of the acoustic sources can be quite complicated representing: compact acoustic sources, diffusive large-scale sources, groups of distant compact sources, or combination of these three types. This reflects the complexity of physical mechanisms responsible for generation of sunquake waves, and flare energy release. In this work, we will not discuss in details this classification because it requires further studies of individual events to develop more clear quantitative criteria.

\begin{small}
\begin{longtable}{|c|c|c|c|c|c|c|c|c|c|}
	
	\caption{Catalog of sunquake events and acoustic sources. The first five columns show general information about the flare: flare start time (UT), GOES class, and standard Active Region number from the NOAA database (AR NOAA), angular distance from the disk center. Last three columns present information about seismic transients during the corresponding solar flares. Plus in column ``SQ movie method'' means presence or absence of sunquake waves in the movies made from the frequency filtered running time differences of the HMI Dopplergrams. Plus in column 'TD method' means that me found the characteristic sunquake wave pattern in the Time-Distance diagram. Plus in column ``Holography method'' means presence or absence of statistically significant acoustic sources determined by the acoustic holography method. Question mark ``?'' indicates events as sunquake candidates, which means that there was a weak acoustic signal around the velocity impact sites during the flare, but the acoustic power was below the three-sigma background noise level. The next three columns  $|H_+|^2$, $S_{max}$ and $df/dt$(1-8 \AA{}) show information about total sunquake energy, area of acoustic sources and maximum values  of the GOES SXR flux time derivative (Energy release rate). The online version of this catalog is available at  \href{https://solarflare.njit.edu/sunquakes/sunquakes.html}{https://solarflare.njit.edu/sunquakes/sunquakes.html}.}

	\label{table1}\\
	\hline
	Date  & GOES   & AR & Location & Movie   & Holog.   & TD & $|H_+|^2$ & $S_{max}$ & $df/dt$(1-8 \AA{}) \\
	UT time & class  &  NOAA  & arcsec & method & method & method & $10^{26}$ & $10^{17}$ & $10^{-7}$  \\
	& & & & & & & ergs & cm$^2$ & Watt m$^{-2}$s$^{-1}$ \\
	\hline
	13.02.2011 17:28	& M6.6 &	11158 &	211 &	  &	? &	  & 3.0 & 4.4 & 5.8 \\
	14.02.2011 17:20	& M2.2 &	11158 &	278 &	  &	? &	  & 3.4 & 2.9 & 3.8 \\
	15.02.2011 01:44    & X2.2 &	11158 &	319 &	+ &	+ &	+ & 10.6 & 9.6 & 10.5 \\
	18.02.2011 09:55    & M6.6 &	11158 &	809 &	+ &	+ &	+ & 12.4 & 9.0 & 12.2 \\
	18.02.2011 12:59	& M1.4 &	11158 &	808 &	  &	+ &	+ & 4.2 & 3.7 & 2.1 \\
	14.03.2011 19:30	& M4.2 &	11169 &	733 &	  &	+ &	  & 12.4 & 11.1 & 7.6 \\
	15.03.2011 00:18    & M1.0 &	11169 &	761 &	  &	? &	  & 6.3 & 4.9 & 3.0  \\
	30.07.2011 02:04    & M9.3 &	11261 &	493 &	+ &	+ &	+ & 33.2 & 15.0 & 14.8 \\
	06.09.2011 22:12	& X2.1 &	11283 &	299 &	  &	? &	  & 0.6 & 0.8 & 27.4   \\
	07.09.2011 22:32	& X1.8 &	11283 &	496 &	+ &	+ &	+ & 9.9 & 9.1 & 24.8 \\
	09.09.2011 12:39	& M1.2 &	11283 &	708 &	  &	? &	  & 5.1 & 6.6 & 1.1   \\
	24.09.2011 20:29	& M5.8 &	11302 &	709 &	  &	+ &	+ & 5.6 & 5.9 & 6.3 \\
	25.09.2011 02:27    & M4.4 &	11302 &	676 &	  &	? &	+ & 3.5 & 4.9 & 3.9 \\
	25.09.2011 08:46    & M3.1 &	11302 &	640 &	+ &	+ &	+ & 4.0 & 5.6 & 4.2 \\
	26.09.2011 05:06    & M4.0 &	11302 &	526 &	+ &	+ &	+ & 14.7 & 12.3 & 6.0 \\
	02.10.2011 17:19	& M1.3 &	11302 &	749 &	+ &	+ &	+ & 18.8 & 6.8 & 3.1 \\
	03.11.2011 20:16	& X1.9 &	11339 &	842 &	+ &	+ &	+ & 27.1 & 19.3 & 19.8 \\
	05.11.2011 20:31	& M1.8 &	11339 &	577 &	  &	+ &	  & 8.7 & 7.4 & 2.7 \\
	25.12.2011 18:11	& M4.0 &	11387 &	462 &	  &	? &	+ & 4.2 & 3.0 & 6.6 \\
	30.12.2011 3:03	    & M1.2 &	11389 &	812 &	  &	+ &	+ & 6.4 & 4.1 & 1.6 \\
	31.12.2011 13:09	& M2.4 &	11389 &	607 &	  &	? &	  & 1.1 & 1.1 & 4.9 \\
	05.03.2012 19:10	& M2.1 &	11429 &	724 &	  &	+ &	  & 6.0 & 12.2 & 3.7 \\
	05.03.2012 19:27	& M1.8 &	11429 &	724 &	  &	+ &	+ & 1.5 & 5.4 & 3.0 \\
	05.03.2012 22:26	& M1.3 &	11429 &	705 &	  &	? &	  & 1.7 & 4.6 & 2.0 \\
	06.03.2012 04:01    & M1.0 &	11429 &	674 &	  &	+ &	+ & 0.5 & 2.0 & 0.8 \\
	06.03.2012 07:52    & M1.0 &	11429 &	654 &	  &	+ &	+ & 2.0 & 5.3 & 0.9 \\
	06.03.2012 12:23	& M2.1 &	11429 &	643 &	  &	+ &	+ & 0.6 & 1.6 & 1.5 \\
	07.03.2012 00:02    & X5.4 &	11429 &	570 &	  &	+ &	+ & 4.6 & 9.1 & 13.7 \\
	09.03.2012 03:22    & M6.3 &	11429 &	380 &	+ &	+ &	+ & 14.5 & 10.3 & 2.7 \\
	08.05.2012 13:02	& M1.4 &	11476 &	636 &	  &	+ &	+ & 1.7 & 3.3 & 1.8 \\
	09.05.2012 12:21	& M4.7 &	11476 &	466 &	  &	? &	+ & 14.7 & 7.7 & 3.0 \\
	09.05.2012 14:02	& M1.8 &	11476 &	452 &	  &	? &	  & 1.0 & 4.0 & 1.7 \\
	09.05.2012 21:01	& M4.1 &	11476 &	397 &	+ &	+ &	+ & 10.2 & 7.3 & 5.3 \\
	10.05.2012 04:11    & M5.7 &	11476 &	341 &	+ &	+ &	+ & 10.0 & 10.1 & 7.7 \\
	10.05.2012 20:20	& M1.7 &	11476 &	246 &	  &	? &	+ & 6.1 & 4.7 & 1.5 \\
	04.07.2012 09:47    & M5.3 &	11515 &	426 &	+ &	+ &	+ & 11.7 & 8.7 & 5.9 \\
	04.07.2012 12:07	& M2.3 &	11515 &	446 &	+ &	  &	+ &  &  &   \\
	04.07.2012 14:35	& M1.3 &	11515 &	454 &	+ &	+ &	+ & 5.1 & 7.3 & 1.7 \\
	05.07.2012 01:05    & M2.4 &	11515 &	509 &	+ &	+ &	+ & 3.1 & 4.7 & 1.6 \\
	05.07.2012 03:25    & M4.7 &	11515 &	519 &	+ &	+ &	+ & 10.2 & 12.9 & 12.5 \\
	05.07.2012 06:49    & M1.1 &	11515 &	538 &	+ &	  &	+ &  &  &   \\
	05.07.2012 10:44	& M1.8 &	11515 &	558 &	+ &	+ &	+ & 1.3 & 4.2 & 2.4 \\
	05.07.2012 11:39	& M6.1 &	11515 &	568 &	+ &	+ &	+ & 20.4 & 19.7 & 10.9 \\
	05.07.2012 20:09	& M1.6 &	11515 &	608 &	+ &	+ &	+ & 9.3 & 6.6 & 3.9 \\
	06.07.2012 01:37    & M2.9 &	11515 &	639 &	+ &	+ &	+ & 9.6 & 9.6 & 4.1 \\
	06.07.2012 13:26	& M1.2 &	11515 &	721 &	+ &	+ &	+ & 2.6 & 2.9 & 2.4 \\
	06.07.2012 23:01	& X1.1 &	11515 &	765 &	+ &	+ &	+ & 8.1 & 7.3 & 11.6 \\
	07.07.2012 03:10    & M1.2 &	11515 &	782 &	  &	+ &	  & 4.0 & 6.6 & 0.8 \\
	07.07.2012 10:57	& M2.6 &	11515 &	820 &	  &	? &	  & 2.3 & 6.2 & 2.8 \\
	09.07.2012 23:03	& M1.1 &	11520 &	610 &	+ &	+ &	+ & 3.1 & 7.8 & 3.7 \\
	23.10.2012 03:13    & X1.8 &	11598 &	789 &	+ &	+ &	+ & 47.1 & 18.5 & 27.7 \\
	13.01.2013 00:45    & M1.0 &	11652 &	478 &	  &	+ &	  & 3.3 & 3.7 & 2.5 \\
	13.01.2013 08:35    & M1.7 &	11652 &	508 &	  &	+ &	+ & 4.0 & 2.5 & 3.9 \\
	17.02.2013 15:45	& M1.9\footnote{The flare of February 17, 2013, is classified as M1.9 GOES-class flare according to the strongest secondary X-ray peak. However, the sunquake event was associated with the first C7.0 peak \citep{Sharykin2015a}} &	11675 &	473 &	+ &	+ &	+ & 3.6 & 2.7 & 10.1 \\
	22.04.2013 10:22	& M1.0 &	11726 &	478 &	  &	? &	  & 1.7 & 3.4 & 2.2 \\
	17.08.2013 18:16	& M3.3 &	11818 &	505 &	  &	? &	  & 2.7 & 2.8 & 3.4 \\
	24.10.2013 10:30	& M3.5 &	11875 &	285 &	+ &	+ &	+ & 3.7 & 3.6 & 5.3 \\
	03.11.2013 05:16    & M5.0 &	11884 &	319 &	  &	? &	  & 7.4 & 5.1 & 6.5 \\
	05.11.2013 18:08	& M1.0 &	11890 &	625 &	+ &	? & + & 0.02 & 1.6 & 1.3 \\
	05.11.2013 22:07	& X3.3 &	11890 &	600 &	+ &	+ &	+ & 31.7 & 17.6 & 70.4 \\
	06.11.2013 13:39	& M3.8 &	11890 &	481 &	+ &	+ &	+ & 12.4 & 11.6 & 4.4 \\
	07.11.2013 03:34    & M2.3 &	11890 &	385 &	+ &	+ &	+ & 6.1 & 6.7 & 2.0 \\
	07.11.2013 14:15	& M2.4 &	11890 &	304 &	+ &	? &	+ & 4.7 & 4.0 & 2.5 \\
	08.11.2013 04:20    & X1.1 &	11890 &	212 &	+ &	+ &	+ & 21.4 & 14.9 & 21.5 \\
	10.11.2013 05:08    & X1.1 &	11890 &	328 &	+ &	+ &	+ & 31.1 & 20.2 & 12.7 \\
	17.11.2013 05:06    & M1.0 &	11900 &	703 &	  &	? &	  & 5.1 & 2.8 & 2.4 \\
	22.12.2013 15:06	& M3.3 &	11928 &	807 &	  &	+ &	  & 10.4 & 3.9 & 2.1 \\
	07.01.2014 10:07	& M7.2 &	11944 &	93  &	+ &	+ &	  & 8.6 & 10.4 & 12.5 \\
	30.01.2014 06:33    & M2.1 &	11967 &	750 &	  &	? &	  & 1.3 & 2.1 & 2.3 \\
	02.02.2014 06:24    & M2.6 &	11968 &	342 &	+ &	+ &	+ & 7.4 & 4.2 & 2.6 \\
	04.02.2014 03:57    & M5.2 &	11967 &	248 &	+ &	+ &	+ & 7.9 & 9.2 & 7.4 \\
	07.02.2014 10:25	& M1.9 &	11968 &	799 &	+ &	+ &	+ & 17.0 & 8.3 & 8.2 \\
	16.02.2014 09:20    & M1.1 &	11977 &	109 &	+ &	+ &	+ & 3.4 & 2.6 & 2.1 \\
	29.03.2014 17:35	& X1.0 &	12017 &	525 &	  &	+ &	  & 5.7 & 5.5 & 13.7 \\
	08.05.2014 09:59    & M5.2 &	12056 &	728 &	+ &	+ &	+ & 2.8 & 5.1 & 2.9 \\
	11.06.2014 05:30    & M1.8 &	12080 &	633 &	  &	? &	  & 6.5 & 6.3 & 2.6 \\
	20.10.2014 09:00    & M3.9 &	12192 &	626 &	  &	? &	+ & 0.5 & 4.3 & 2.1 \\
	20.10.2014 18:55	& M1.4 &	12192 &	553 &	+ &	? &	  & 10.5 & 7.0 & 2.3 \\
	22.10.2014 01:16    & M8.7 &	12192 &	388 &	+ &	+ &	+ & 5.4 & 16.1 & 2.1 \\
	22.10.2014 14:02	& X1.6 &	12192 &	290 &	+ &	+ &	+ & 1.7 & 5.5 & 4.3 \\
	23.10.2014 09:44    & M1.1 &	12192 &	226 &	  &	+ &	+ & 1.9 & 7.3 & 1.1 \\
	26.10.2014 18:07	& M4.2 &	12192 &	667 &   + &	+ &	+ & 3.0 & 8.3 & 3.0 \\
	09.11.2014 15:24	& M2.3 &	12205 &	297 &	  &	+ &	+ & 8.7 & 8.7 & 2.4 \\
	15.11.2014 20:38	& M3.7 &	12209 &	713 &	  &	? &	  & 3.6 & 3.7 & 5.0 \\
	20.12.2014 00:11    & X1.8 &	12242 &	502 &	+ &	? &	+ & 3.5 & 12.6 & 6.6 \\
	03.01.2015 09:40    & M1.1 &	12253 &	267 &   + &	+ &	+ & 4.3 & 5.6 & 1.6 \\
	30.01.2015 12:10	& M2.4 &	12277 &	797 &	  &	+ &	  & 9.2 & 7.8 & 2.4 \\
	10.03.2015 03:19    & M5.1 &	12297 &	588 &	+ &	+ &	+ & 12.5 & 11.2 & 7.7 \\
	10.03.2015 23:46	& M2.9 &	12297 &	442 &	  &	+ &	  & 8.1 & 8.0 & 5.1 \\
	11.03.2015 16:11	& X2.1 &	12297 &	333 &	  &	+ &	+ & 2.6 & 4.6 & 12.2 \\
	12.03.2015 04:41    & M3.2 &	12297 &	245 &	  &	+ &	+ & 4.3 & 5.4 & 4.5 \\
	12.03.2015 13:50	& M4.2 &	12297 &	186 &	  &	? &	+ & 3.1 & 3.5 & 2.3 \\
	12.03.2015 21:44	& M2.7 &	12297 &	168 &	  &	? &	  & 3.5 & 4.8 & 3.4 \\
	15.03.2015 09:36    & M1.0 &	12297 &	502 &	+ &	+ &	+ & 3.6 & 6.1 & - \\
	21.06.2015 09:38    & M3.8 &	12367 &	807 &	  &	? &	  & 7.4 & 5.4 & 6.6 \\
	25.06.2015 08:02    & M7.9 &	12371 &	653 &	+ &	+ &	+ & 7.6 & 8.7 & 13.1 \\
	22.08.2015 21:19	& M3.5 &	12403 &	376 &	+ &	+ &	+ & 18.1 & 14.3 & 5.1 \\
	24.08.2015 07:26    & M5.6 &	12403 &	377 &	  &	? &	  & 5.3 & 5.7 & 8.9 \\
	24.08.2015 17:40	& M1.0 &	12403 &	417 &	  &	+ &	  & 5.3 & 5.0 & 2.7 \\
	28.09.2015 14:53	& M7.6 &	12422 &	535 &	+ &	+ &	+ & 20.5 & 15.9 & 10.7 \\
	29.09.2015 06:39    & M1.4 &	12422 &	622 &	+ &	  &	+ &  &  &   \\
	30.09.2015 13:18	& M1.1 &	12422 &	766 &	+ &	+ &	+ & 8.3 & 4.1 & 3.9 \\
	01.10.2015 13:03	& M4.5 &	12422 &	866 &	  &	+ &	  & 3.6 & 3.3 & 4.1 \\
	31.10.2015 17:48	& M1.0 &	12443 &	698 &	  &	+ &	  & 4.5 & 5.4 & 2.8 \\
	04.09.2017 15:11	& M1.5 &	12673 &	294 &	  &	? &	  & 6.9 & 9.9 & 1.7 \\
	04.09.2017 20:28	& M5.5 &	12673 &	341 &	+ &	+ &	+ & 2.2 & 5.0 & 4.6 \\
	05.09.2017 01:03    & M4.2 &	12673 &	349 &	+ &	+ &	+ & 2.5 & 4.6 & 6.2 \\
	06.09.2017 08:57    & X2.2 &	12673 &	575 &	+ &	+ &	+ & 9.0 & 28.2 & - \\
	06.09.2017 11:53	& X9.3 &	12673 &	599 &	+ &	+ &	+ & 58.9 & 42.4 & 63.1 \\
	07.09.2017 04:59    & M2.4 &	12673 &	694 &	+ &	+ &	+ & 9.2 & 13.2 & 5.4 \\
	07.09.2017 10:11	& M7.3 &	12673 &	751 &	+ &	+ &	+ & 24.9 & 12.8 & 33.5 \\
	07.09.2017 14:20	& X1.3 &	12673 &	747 &	+ &	+ &	+ & 24.7 & 23.6 & 14.3 \\
	08.09.2017 02:19    & M1.3 &	12673 &	788 &	+ &	+ &	+ & 4.9 & 9.2 & 2.9 \\
	08.09.2017 07:40    & M8.1 &	12673 &	816 &	+ &	+ &	+ &  9.9 & 9.6 & 7.9 \\
	\hline
\end{longtable}
\end{small}



One of the most interesting findings is that we were able to distinguish active regions which gave the largest contribution to the total number of sunquakes. In Table~\ref{table2} we present a summary of the most ``helioseismically efficient'' active regions which produced five or more sunquakes. All these active regions had a complex magnetic structure, and were characterized by the Hale class of $\beta\gamma\delta$. The largest number of sunquakes was generated in AR~12673 during September 4-7, 2017. It is also worth mentioning that this active region produced the strongest sunquake (ever observed during Cycle 24) during the X9.3 solar flare of September 6, 2017, 12:53 UT \citep[][]{Sharykin2018}. We define a relative seismic efficiency as ratio $N_{1}/N_{MX}$, where $N_{MX}$ is the total number of M and X solar flares in an active region, observed on the disk, and $N_{1}$ is the total number of sunquake events detected at least by one method (without candidates). According to this definition, AR~11890 was the most seismically efficient.

\begin{table}[h]
\caption{Characteristics of active regions that produced five or more sunquakes. $N_{MX}$ is the total number of M and X on-disk solar flares in the active regions. $N_{dV}$ is the total number of flares with Doppler velocity impacts. $N_{SQ}$, $N_{TD}$ and $N_{AH}$ are the numbers of sunquake events detected by using the sunquake movie, the Time-Distance diagram and the Acoustic Holography methods, respectively. $N_{All}$ is the total number of sunquake events detected by all three methods, and $N_{1}$ is the number of sunquakes detected by at least by one method. Ratio $N_{1}/N_{MX}$ defines the seismic efficiency of AR. The last two rows summarize the sunquake efficiency numbers.}
\begin{tabular}{|c|c|c|c|c|c|c|c|c|}     
  \hline                   
NOAA AR & $N_{MX}$ & $N_{dV}$ & $N_{SQ}$ & $N_{TD}$ & $N_{AH}$ & $N_{All}$ & $N_{1}$ &  $N_{1}/N_{MX}$  \\
  \hline
 11302 & 19(1) & 7  & 3  & 5  & 4(1)  & 3  & 5  & 0.26 \\
 11429 & 15(1) & 11 & 1  & 6  & 7(1)  & 1  & 7  & 0.47 \\
 11476 & 6(0)  & 6  & 2  & 5  & 3(3)  & 2  & 5  & 0.83 \\
 11515 & 27(1) & 17 & 12 & 12 & 11(1) & 10 & 13 & 0.48 \\
 11890 & 8(3)  & 8  & 7  & 7  & 5(2)  & 5  & 7  & 0.88 \\
 12192 & 26(5) & 12 & 4  & 5  & 4(2)  & 3  & 6  & 0.23 \\
 12297 & 19(1) & 9  & 2  & 5  & 5(2)  & 2  & 6  & 0.32 \\
 12673 & 24(3) & 18 & 9  & 9  & 9(1)  & 9  & 9  & 0.38 \\
  \hline
 Total & 144(15) & 88 & 40 & 54 & 48(13) & 35 & 58 & \\
   \hline
 Total in catalog & 181(25) & 181 & 62 & 80 & 81(29) & 54 & 94 & \\
  \hline
\end{tabular}
\label{table2}
\end{table}

\section{RESULTS OF STATISTICAL ANALYSIS OF SUNQUAKE AND GOES X-RAY DATA}

The presented catalog opens opportunities for statistical studies of the sunquake phenomenon. In this paper, we present initial results of our investigation of the relationship between the acoustic energy of sunquakes and the flare characteristic derived from the soft X-ray (SXR) data obtained from the GOES satellite \citep{Bornmann1996}. In addition, we compare the X-ray characteristics of flares that produced helioseismic signals and flares without photospheric perturbations.
In this study, we do not consider potential candidates.

We use the following formula to estimate the total sunquake power $H_{+}$[erg/s] in the frequency range 5-7 mHz:
$$
H_{+} = \int_{S_{ROI}}\int_{t_1}^{t_2}c_s\frac{\rho\delta v^2}{2}dSdt.
$$
Here $S_{ROI}$ means the area of regions of interest where we observed photospheric perturbations and acoustic sources deduced by the holography technique; $t_{st}$ and $t_{fn}$ are times of the onset and end of a sunquake event, defined as the time moments when the total acoustic power is above three-sigma level of the background noise (see examples in Figures~\ref{Egress1}a and~\ref{Egress2}a). The background noise is determined as the spatially and temporary  averaged value of the acoustic preflare power in the region of interest (ROI shown by dashed contours in Figures~\ref{Egress1}b-g and~\ref{Egress2}b-g). The acoustic energy flux is calculated as $c_s\rho\delta v^2/2$, where $c_s$ is the photospheric sound speed, and $\delta v$ is the amplitude of acoustic perturbations.

\begin{figure}[ht]
\centering
\includegraphics[width=1.0\linewidth]{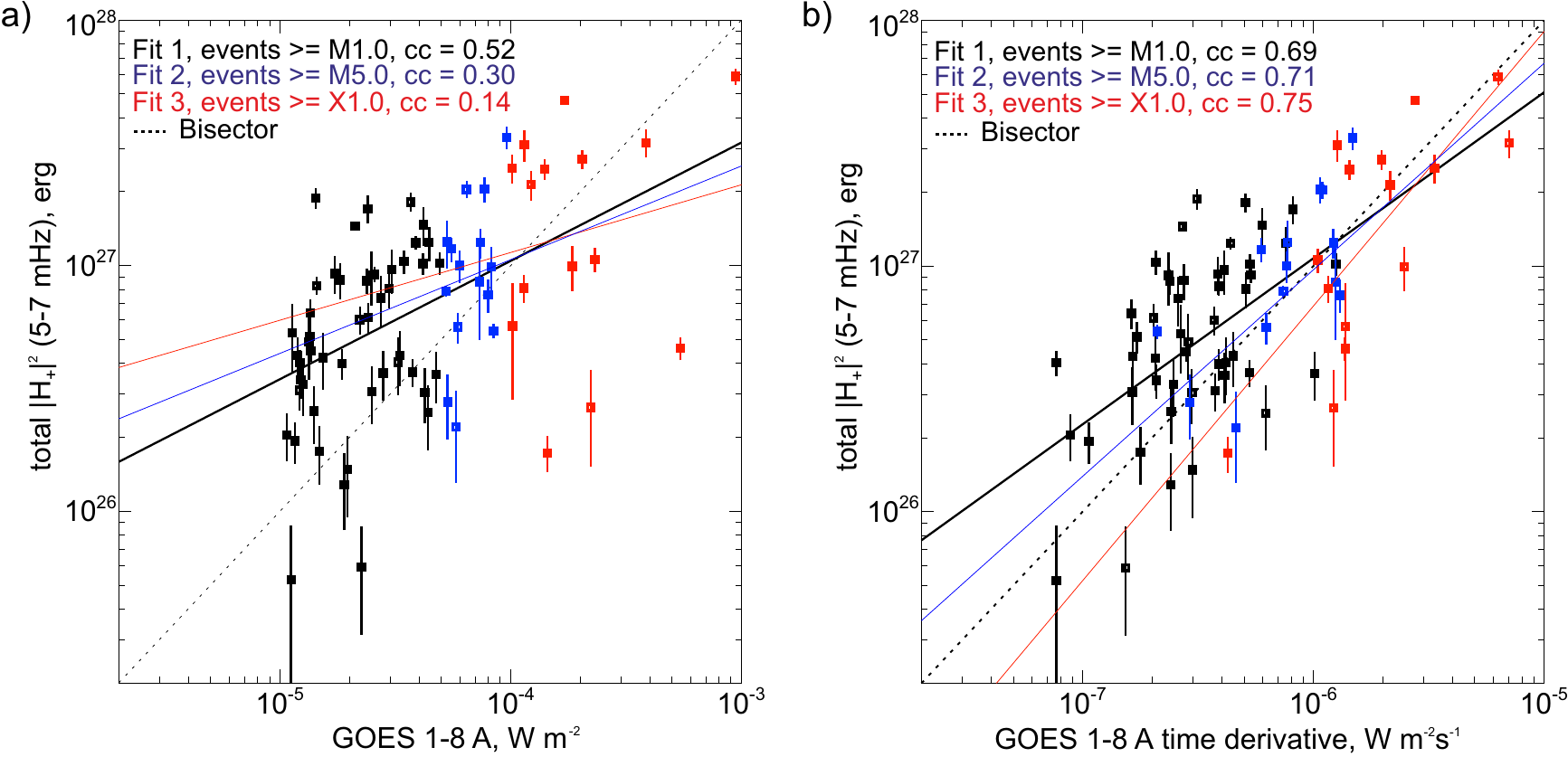}
\caption{Relationship between the total acoustic energy ($|H_+|^2$) in the frequency range 5-7 mHz and the Soft X-ray (SXR) flux (panel~a), and maximum value of the SXR time derivative (panel b). Linear correlation coefficients are written within panels for three flare classes: 1) flares with the SXR class higher than M1.0, 2) higher than M5.0 and 3) only X-class events. Bisector dashed line is also plotted.}
\label{Stat1}
\end{figure}

Figure~\ref{Stat1} presents two statistical plots: (a) comparison of the total sunquake power with the corresponding maximum value of the GOES SXR flux time derivative, and (b) comparison with the maximum value of the GOES SXR flux in the wavelength band of 1-8~\AA{}. We show the error bars for the sunquake power considering only the acoustic background. Potential systematic uncertainties related to deviations of the solar interior structure from the standard model are not taken into account in these estimates. Colors in Figure~\ref{Stat1}  highlight flares of the different GOES class ranges: M- and X-class flares (black), higher than M5.0 (blue) and X-class flares (red). These plots reveal positive linear correlations for both comparisons. However, the maximum value of the SXR flux time derivative shows significantly better correlation with the sunquake energy (the correlation coefficient is 0.69-0.75 for the different GOES class ranges) than with the peak SXR value(the correlation coefficient is 0.14-0.52). It means that the total helioseismic energy is mostly related to the flare-energy release rate. In other words, flares that are more impulsive are also more seismic.

\begin{figure}[ht]
\centering
\includegraphics[width=\linewidth]{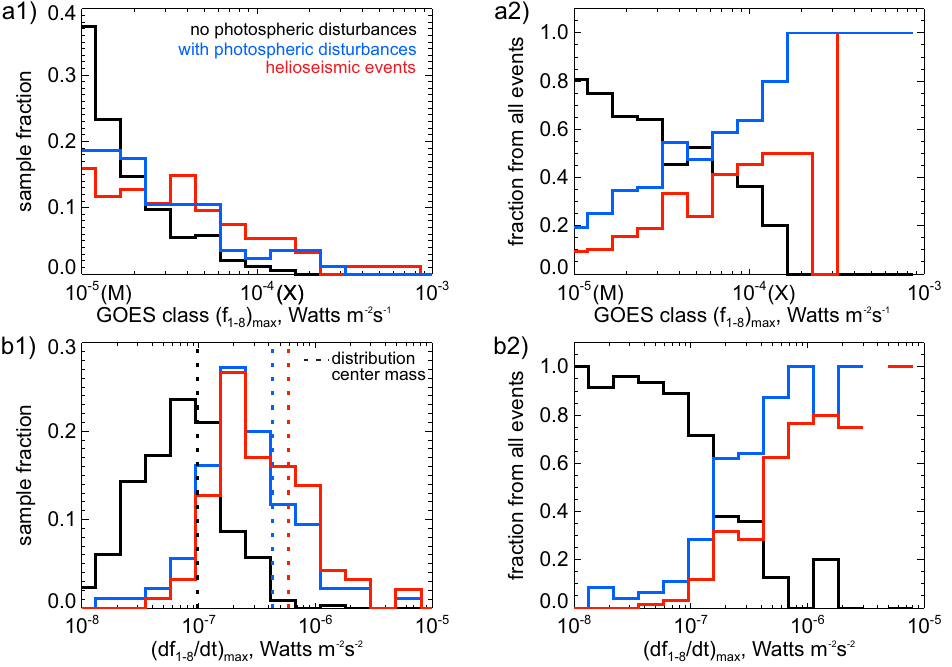}
\caption{Histograms for three types of flares: 1)  without photospheric perturbations (black lines), 2)  with photospheric perturbations (blue) including sunquakes, 3) with sunquakes (red). Left column shows relationship between event occurrence within the types and the SXR flux (a1) and the maximum value of the SXR flux time derivative (b1). Right column shows probability functions that determine the event occurrence relative to the total number of all flares for the SXR flux (a2) and the maximum value of the SXR flux time derivative (b2).}
\label{hist1}
\end{figure}
\begin{figure}[ht]
\centering
\includegraphics[width=0.95\linewidth]{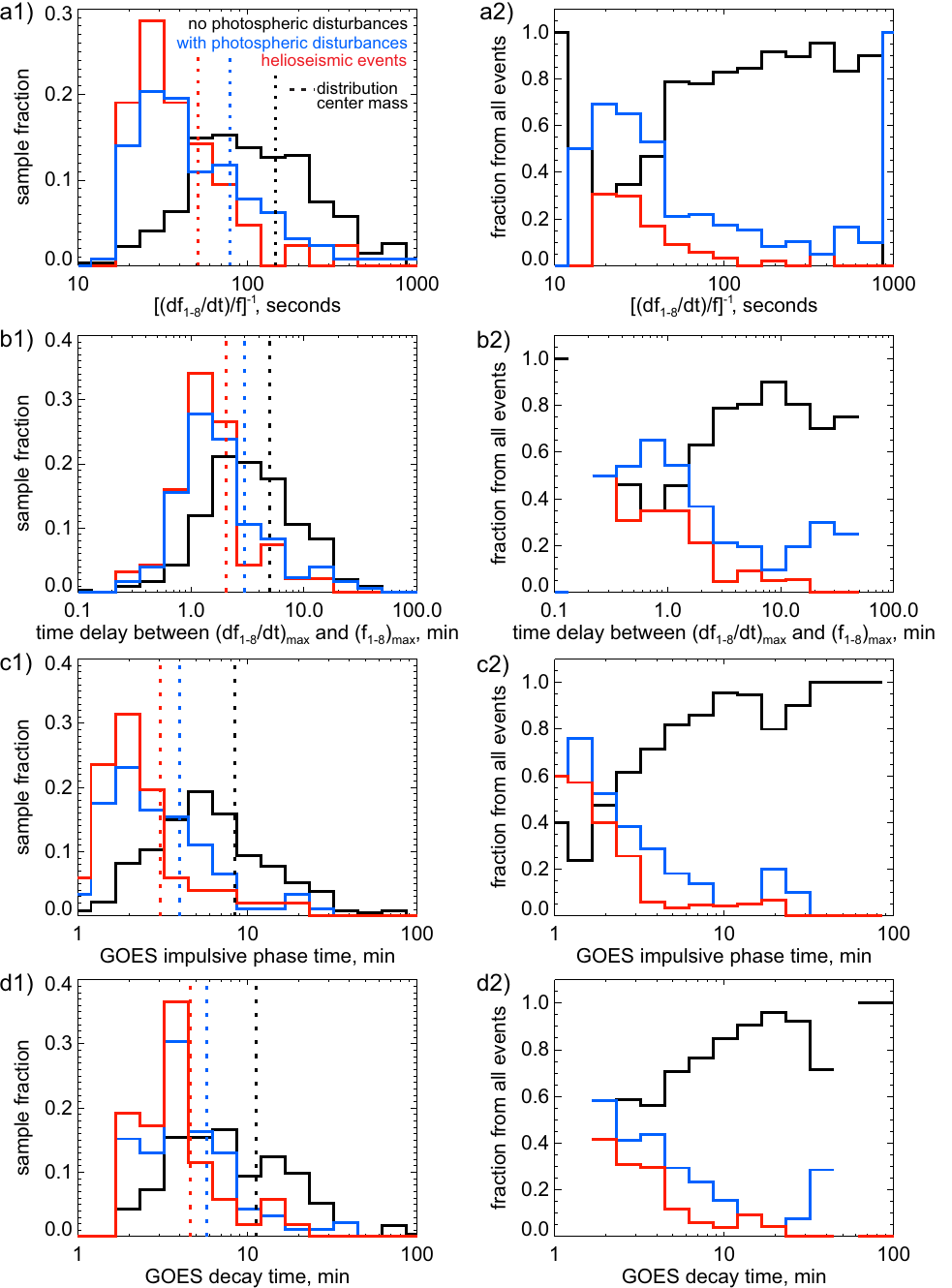}
\caption{Histograms for the same three types of flares as in Fig.~\ref{hist1} for: a1-a2) the characteristic energy release time estimated as the maximum value of $f_{1-8}/(df_{1-8}/dt)$; b1-b2) for the time delays between the peak times of $f_{1-8}$ and $df_{1-8}/dt$; c1-c2) the impulsive phase duration; d1-d2) the flare decay time.
}
\label{hist2}
\end{figure}

To compare properties of the flares with and without sunquake events, in Figures~\ref{hist1} and~\ref{hist2} we plot histograms for three groups:  1) seismically active solar flares (red); 2) flares with photospheric perturbations including sunquakes (blue); 3) flares without photospheric impacts (black).
Figure~\ref{hist1} shows the flare distributions vs the flare maximal SXR fluxes (panels $a1-a2$), and vs the maximum value of the SXR flux time derivative, which characterizes the maximal energy release rate (panels $b1-b2$). Left columns show the distributions of events in each of the three groups normalized to the total number of events in these groups as a function of the selected parameters (shown in X-axis). The right column shows the occurrence rate of events within the three groups relative to the total number of flares (with and without photospheric perturbations). In other words, the panels in right columns show probability functions. Value 1 means that all flares in the particular range $\Delta X_i$ of a parameter represented by the X-axis are sunquakes, photospheric (including sunquakes) events, or events without sunquakes (indicated by red, blue, black color respectively).

These distributions show that the appearance of helioseismic waves is more probable for the flare of higher GOES X-ray classes with higher energy release rates. We can introduce a formal criterion for appearance of the flare helioseismic response: with probability higher than 60~\% sunquakes will be registered when the maximal flare energy release rate is higher than $2\times 10^{-7}$~Watts~m$^{-2}$s$^{-1}$. 

Comparing the centers of mass calculated for the distributions (dashed lines in Fig.~\ref{hist1}b1), we determine that the seismic flares are 6 times more impulsive than the flares without photospheric impacts. If we consider all flares with photospheric perturbations (blue histograms) then we see only small differences compared to the histogram for the seismic flares (red lines). Nevertheless, we can state that the flares with photospheric impacts but without pronounced helioseismic waves are less impulsive but still have high-energy release rates than the flares without photospheric perturbations.

This leads us to a conclusion that the seismic flares are more impulsive than the non-seismic flares in terms of the maximal flare-energy release rate. 
To confirm the impulsive nature of the flares producing sunquakes, we made an additional comparative analysis of the three groups from the point of view of the characteristic energy release times.

Figure~\ref{hist2} presents distributions of the two types (similarly to Fig.\,\ref{hist1}, sorted in two columns) illustrating differences between the characteristic flare times of the seismic and non-seismic flares. Panels (a1)-(a2) present the characteristic energy release time estimated as the maximum value of $f_{1-8}/(df_{1-8}/dt)$, where $f_{1-8}$ is the GOES SXR flux in the 1-8~\AA{} channel. Panels (b1)-(b2) show distributions of the time delays between the peak times of $f_{1-8}$ and $df_{1-8}/dt$. Distributions of the impulsive phase duration defined as the time interval when the SXR flux was higher than $max(f_{1-8})/10$ are shown in panels (c1)-(c2). Panels (d1)-(d2) show the flare distribution for the SXR decay time determined as the time interval during which the SXR flux decreased by a factor of two from its maximum. We also calculated the centers of mass of the distributions for easier comparison among the flare classes. From these distributions, we find that the seismic flares are characterized by shorter durations compared to the non-seismic flares. In other words, sunquake events are more probable for flares that are more impulsive. This difference is more pronounced in terms of the maximum values of the SXR flux time derivative.

\section{CONCLUSIONS}

We have performed a comprehensive search for seismically active (producing sunquakes) flares among all M-X class solar flares observed in Solar Cycle 24. Using the new catalog of sunquakes, we performed a comparative statistical analysis of the X-ray emission from the GOES data and its temporal dynamics for seismic and non-seismic flares. The obtained results can be summarized as follows:

\begin{enumerate}

\item We have found that 94 flares among 507 flares of the X-ray class greater than M1.0 were seismically active. Our analysis has shown that there are many solar flares of moderate class with strong sunquakes, while in some powerful X-class flares, helioseismic waves were not observed or were weak.

\item Our analysis also revealed that during Solar Cycle 24, there were several active regions characterized by the most efficient generation of sunquakes.

\item We found that the sunquake total energy correlates with the maximum value of the soft X-ray time derivative better (with correlation coefficient $\approx 0.7$) than with the X-ray class \citep[contrary to what one could expect from the ``big-flare syndrome'' idea,][]{Kahler1982}. In this case, the impulsiveness of the energy release plays an important role.

\item It was shown that the flares producing sunquakes are more impulsive (shorter flare times and higher heating rate) compared to the flares without photospheric perturbations. The most evident difference between distributions of the seismic and non-seismic flares appears in terms of the maximum values of the flare-energy release rate.

\end{enumerate}

In the presented initial statistical study of the new sunquake catalog we investigated the relationship of the sunquake energetics with the temporal and amplitude characteristics of the flare soft X-ray emission. Although the results do not allow us to answer the long-standing question about the exact mechanism of the flare helioseismic response, they show that the impulsiveness of the flare energy release characterized by the SXR flux derivative (the Neupert effect) plays an important role in the sunquake mechanism. The impulsive nature of seismic flares gives hints that they are compact with fast energy release rate. Thus, one can assume that low-lying short magnetic loops are involved into the flare-energy release process. 

In our catalog, the helioseismic events are tagged to the standard GOES classification of solar flares in terms of the peak of soft X-ray emission. Our previous studies have shown that the sunquake initiation may occur at the very beginning of the flare impulsive phase \citep{Sharykin2017}, or during a subflare X-ray peak \citep{Sharykin2015a}. Thus, further detailed studies considering characteristics of the magnetic field structure and flare emission in other ranges of electromagnetic spectrum are needed to shed more light on the sunquake mechanism.

The created catalog of helioseismic solar flares opens new opportunities for performing statistical analyses, as well as for in-depth research of individual cases. It will contribute to better understanding of the mechanism of sunquakes and flares.

\acknowledgments
The research was supported by the NASA Grants NNX14AB68G and NNX16AP05H, NSF grant 1916509, and grant of the President of the Russian Federation for the State Support of Young Russian Science PhDs (MK-5921.2018.2). The research is also partially supported by the Russian Foundation of Basic Research (RFBR, grant 18-02-00507). The observational data are courtesy of the SDO/HMI, and GOES science teams.


\bibliographystyle{apj}

\clearpage



\end{document}